\documentclass[prc,twocolumn,amsmath,amssymb,superscriptaddress,noffotinbib,preprintnumbers]{revtex4-1}
\usepackage{url,color,colordvi,epsfig,pstricks}
\usepackage[colorlinks=true, citecolor=blue, filecolor=blue, linkcolor=blue, urlcolor=blue, pdftex]{hyperref}
\usepackage{graphicx,bm,dcolumn}
\usepackage{hyperref,slashed,lineno}
\usepackage{physics}
\usepackage{inputenc}

\hypersetup{
    colorlinks=true, 
    linktoc=all, 
    linkcolor=blue, 
    citecolor=blue}

\newcommand{\cgc}[6]{ \left\langle #1 \: #2 \; #3 \: #4 \right| #5 \: #6 \left.\right\rangle }

\begin{document}

\title{Benchmarking intra-nuclear cascade models for neutrino scattering with relativistic optical potentials.}
\author{A.~Nikolakopoulos}\email{alexis.nikolakopoulos@ugent.be}
\affiliation{Department of Physics and Astronomy, Ghent University, B-9000 Gent, Belgium}
\affiliation{Theoretical Physics Department, Fermilab, Batavia, IL 60510, USA}
\author{R.~Gonz\'alez-Jim\'enez}
\affiliation{Grupo de F\'isica Nuclear, Departamento de Estructura de la Materia, F\'isica T\'ermica y Electr\'onica, Facultad de Ciencias F\'isicas, Universidad Complutense de Madrid and IPARCOS, CEI Moncloa, Madrid 28040, Spain}
\author{N.~Jachowicz}
\affiliation{Department of Physics and Astronomy, Ghent University, B-9000 Gent, Belgium}
\author{K.~Niewczas}
\affiliation{Department of Physics and Astronomy, Ghent University, B-9000 Gent, Belgium}
\affiliation{University of Wrocław, Institute of Theoretical Physics, Plac Maxa Borna 9, 50-204 Wrocław, Poland}
\author{F.~S\'anchez}
\affiliation{University of Geneva, Section de Physique, DPNC, Geneva, Switzerland}
\author{J.~M.~Ud\'ias}
\affiliation{Grupo de F\'isica Nuclear, Departamento de Estructura de la Materia, F\'isica T\'ermica y Electr\'onica, Facultad de Ciencias F\'isicas, Universidad Complutense de Madrid and IPARCOS, CEI Moncloa, Madrid 28040, Spain}


\begin{abstract}
\begin{description}
\item[Background]  
In neutrino oscillation experiments, the hadrons created in neutrino-nucleus collisions are becoming important  observables.
The description of final-state interactions (FSI) of hadrons with nuclei in the large phase space probed in these experiments poses a great challenge. 
In the analysis of neutrino experiments, which operate under semi-inclusive conditions, cascade models are commonly used for this task. The description of FSI under exclusive conditions on the other hand can be treated successfully by using relativistic optical potentials (ROP).
\item[Purpose] 
We formulate conditions under which the ROP approach and cascade model can be directly compared.
Through this comparison the treatment of FSI in cascade models is studied and benchmarked.
\item[Method] 
We study single proton knockout with the T2K near-detector muon neutrino flux.
We feed the NEUT cascade model with events distributed according to the cross section of a relativistic distorted-wave impulse approximation (RDWIA) calculation that uses the real part of an optical potential (rROP).
We impose cuts on the missing energy of the resulting events to define a set of events which undergo only elastic FSI, these can be compared to RDWIA calculations with the full optical potential.
\item[Results]
The NEUT cascade and ROP give similar cross sections for proton kinetic energies $T_p > 150~\mathrm{MeV}$ for carbon, oxygen and calcium nuclei. A necessary condition is that a realistic nuclear density is used to introduce events in the cascade.
For $T_p < 100~\mathrm{MeV}$ the ROP and NEUT cross sections diverge strongly in shape and differences in magnitude are larger than 50 \%.
Single transverse variables allow to discriminate between different approaches to FSI, in particular the large $\delta \alpha_T$ region is sensitive to the presence of non-elastic FSI. Due to experimental errors and a large non-QE contribution the comparison to T2K data does not give an unambiguous view of FSI.
We discuss electron scattering data and provide results for kinematics covered in the $e4\nu$ analysis.
We argue that with a simple cut in missing energy FSI can be studied with minimal confounding factors.

\item[Conclusions]
 The agreement of the ROP and NEUT cascade under T2K conditions lends confidence to these models as a tool in neutrino oscillation analyses for sufficiently large nucleon kinetic energies.
Our results urge for caution when a cascade model is applied for small nucleon energies. The assessment of model assumptions relevant to this region are strongly encouraged.
The approach presented in this paper provides novel constraints on cascade models from proton-nucleus scattering and can be easily applied to other neutrino event generators.
\end{description}
\end{abstract}

\maketitle


\section{Introduction}
The latest generations of neutrino detectors have the capability to measure (part of) the hadronic final-state in neutrino-nucleus collisions, and oscillation experiments may rely on this information in their analysis~\cite{NUSTECWP,DUNE16,MicroBooNE:2020akw, MicroBooNE:2015bmn,T2K:2018rnz}.
In accelerator-based neutrino experiments the incoming neutrino flux is broad, and the exact neutrino energy is a priori unknown on an event-to-event basis.
For a dependable analysis all reaction channels that contribute to the experimental signal need to be accounted for. 
The problem posed stands in stark contrast to a traditional electron scattering experiment where the incoming energy is known, and the kinematics are selected carefully to study properties of the nucleus with minimal complications.
In an accelerator-based neutrino experiment, on the other hand, the total energy of the hadron system is a priori unknown, and the kinematics of all accessible final-state hadron configurations need to be described over the whole phase space.
The complete description of final-state interactions (FSI) of hadrons with nuclei over the large phase space that is probed poses an unprecedented challenge for nuclear theory.
The complexity depends strongly on the experimental signal that needs to be described, but in general there is no microscopic theory that can deal in a tractable way with this severe coupled-channels problem posed by neutrino experiments.
For this reason, cascade models are used in experimental analyses to model FSI and predict the kinematics and multiplicity of hadrons in the final state. Several cascade models, some of which specifically target neutrino-nucleus scattering in the few-GeV region, are available~\cite{FSIGENIE,NEUT:2021EPJST, NuWro:FSI, Isaacson:cascade2021, Salcedo88}.
The approach that is used in commonly used neutrino event generators such as NEUT~\cite{Hayato:NEUT,NEUT:2021EPJST}, GENIE~\cite{GENIE} and NuWro~\cite{WRONG, NuWro:FSI} is that the initial interaction and the final state cascade are treated as discrete steps.
For one-nucleon knockout, for example, a nuclear model is used for the primary interaction which produces a nucleon with a certain four-momentum. This nucleon is then introduced in the cascade model at some radius, 
and propagated through the cascade to generate the final-state kinematics of the nucleon and other particles that can be detected. 
In many cases, as in Refs.~\cite{Dolan:2019bxf,Dolan:2021rdd}, only the inclusive cross section is explicitly known. The kinematics of the final-state nucleon are determined by selecting an initial nucleon from some momentum distribution and then applying energy and momentum conservation.
This factorization of the process may lead to inconsistency when the outgoing nucleon kinematics are not computed from the same nuclear model as the inclusive cross section. Moreover, the dependence of the cross section on the full set of relevant independent kinematic variables is lost in this way.

Intra-nuclear cascades model the total reaction cross section in nucleon-nucleus collisions by scattering with the constituent nucleons. The nucleon-nucleon cross section can be further broken down into e.g. elastic  and inelastic scattering as in the NEUT cascade~\cite{NEUT:2021EPJST}. Experimental data for total reaction cross sections, as well as the nuclear transparency, are used to constrain and validate these models~\cite{ValidationmethodsINC:2021, Niewczas:Transparency}.
Most of the intra-nuclear cascade models used in neutrino scattering experiments converge to the same cross sections and transparancy for nucleons with sufficient energy, but give different predictions at low energies where the importance of nuclear and quantum mechanical effects increases~\cite{ValidationmethodsINC:2021}.

A more traditional, relativistic and quantum-mechanical approach to nucleon-nucleus scattering comes in the form of (empirical) relativistic optical potentials (ROP).
An empirical potential can be obtained by fitting the angle and energy dependence of the elastic proton-nucleus scattering cross section~\cite{Cooper:1993,Cooper:2009}. 
The imaginary part of the potential `absorbs' all nucleons which undergo inelastic interactions.
The empirical potential then reproduces, in addition to the details of the elastic cross section, the energy-dependence of the total and reaction cross sections, but does not explicitly describe the inelastic interactions.
The effect of FSI in electroweak nucleon-knockout from nuclei, can be modeled in the relativistic distorted wave impulse approximation (RDWIA) by treating the outgoing nucleon as a scattering state in such a ROP.
This approach is used under exclusive conditions, and has been extensively applied to describe electron-induced nucleon knockout $(e,e^\prime p)$ from nuclear shells~\cite{BOFFI1993,Dickhoff:2004,Udiaseep,Udias01, RDWIA:Kelly}.
The ROP is suitable in this case because of the restrictive selection of the missing-energy in such experiments.
Under these conditions, the nucleons that undergo inelastic interactions are not part of the experimental signal and the nucleon flux lost to these inelastic channels is removed by the imaginary part of the potential.
This means that a calculation with an optical potential will underestimate the inclusive cross section, for which all the flux lost in inelastic channels has to be retained.
To describe the inclusive cross section then, an approach like the relativistic Green's function method of Refs.~\cite{Meucci03,Giusti2012MB,Boffi93} can be used.
A simpler treatment is to retain only the energy-dependent real part of the optical potential (rROP) in the RDWIA to describe the inclusive cross section. This rROP approach is found to be very effective in describing the inclusive electroweak cross section~\cite{Gonzalez-Jimenez:2019ejf, Gonzalez-Jimenez19, Kim03}.

In this work, we perform a consistent comparison of the FSI treatment in the NEUT intra-nuclear cascade model with the  microscopic RDWIA approach that uses empirical ROP.
To ensure consistency between NEUT and the RDWIA, and to avoid the problem of factorization as described above, we supply the cascade model with events generated from an unfactorized RDWIA calculation that uses the rROP.
In this way, the nuclear model used as input of the cascade is the same as that in the ROP. Moreover, after summation and integration over all final-state configurations that result from the cascade a realistic inclusive cross section is recovered.
We then apply a cut on missing energy to the events resulting from the cascade model to obtain a sample in which events that undergo inelastic FSI are removed.
The resulting cross sections obtained from the NEUT cascade are directly comparable to RDWIA calculations that use the full ROP.
Through this direct comparison, the treatment of FSI in cascade models is isolated and benchmarked with the well-established microscopic RDWIA approach. 
This work presents a method through which the extensively studied phenomenology of elastic proton-nucleus scattering can be used to provide novel constraints for any nuclear cascade model used in the analysis of neutrino experiments.

This paper is structured as follows: In section~\ref{sec:formalism} we describe the RDWIA approach and discuss the different potentials that are used to treat FSI. In section~\ref{sec:T2KTP} we compare the distribution of proton energy from the NEUT cascade when supplied with RDWIA events that are able to reproduce the inclusive cross section with the optical potential treatment. In subsections~\ref{sec:rdep} and~\ref{sec:Adep} we further examine the influence of the nuclear density and the A-dependence of the results.
Then, in section~\ref{sec:T2Kdata}, we confront the results to T2K data.
Finally we discuss what insight can be gained from electron scattering data, and provide results for the kinematics available in the $e4\nu$ analysis of CLAS data in section~\ref{sec:electron}.
Conclusions and prospects are given in section~\ref{sec:conclusion}.

\section{Formalism \label{sec:formalism}}
We consider the neutrino ($\nu_l$) induced charged-current one-proton ($p$) knockout process
\begin{equation}
    \label{eq:FVproc}
    \nu_l(k_i) + A(P_i) \rightarrow l(k_f) + p(k_N) + B(P_B),
\end{equation}
where $A$ is the initial state nucleus, $B$ is the residual hadron system which remains undetected and the symbols between brackets denote the respective particle's four-vector e.g. $k_i = (E_i=T_i+M_i,\vec{k}_i)$.
With the direction of the incoming beam fixed and the target nucleus at rest all four-vectors in Eq.~(\ref{eq:FVproc}) are fully determined by 7 independent kinematic variables. These may be chosen as $(\lvert \vec{k}_f \rvert, \Omega_f, \lvert \vec{k}_N \rvert, \Omega_N, E_m)$, with $\Omega$ the particle's solid angle, where an overall azimuth angle (e.g. $\phi_f$) is trivial.
The missing energy $E_m$ relates the incoming energy and the kinetic energy of the residual system as 
\begin{equation}
\label{eq:EM}
    E_m  = E_i - E_f - T_N - T_B = M_B + M_N - M_A.
\end{equation}
The total energy of the residual system is
\begin{equation}
    \label{eq:EB}
    E_B = T_B + M_B = \sqrt{M_B^2 + \lvert \vec{p}_{m} \rvert^2},
\end{equation}
and its momentum is the missing momentum
\begin{equation}
    \label{eq:PM}
    \vec{p}_m = \vec{k}_i - \vec{k}_f - \vec{k}_N.
\end{equation}
The probed values of missing energy depend on the processes that are considered (either explicitly or implicitly) for the composition of the residual system $B$ and the nuclear model that is used.

In a neutrino experiment the incoming energy distribution is described by a broad flux $\Phi(E_i)$, and as such, under the assumption that apart from $\vec{k}_f$ and $\vec{k}_N$ the final state particle content is not known, the missing energy cannot be measured on an event-to-event basis. The probability of finding a $1l 1p$ event is proportional to the flux averaged cross section
\begin{align}
&\left<\frac{\,d^5\sigma}{\,d \lvert \vec{k}_f\rvert \,d\cos\theta_f \,d \lvert \vec{k}_N \rvert \,d\Omega_N} \right> \nonumber  \\ &= \int \,d E_m \tilde{\Phi}(E_i) \frac{\,d^5\sigma (E_m)}{\,d \lvert \vec{k}_f\rvert \,d\cos\theta_f \,d \lvert \vec{k}_N \rvert \,d\Omega_N},
\end{align}
where $\tilde{\Phi}(E_\nu)$ is the normalized neutrino flux, with $E_i$ given by Eq.~(\ref{eq:EM}).

We will use the relativistic mean field (RMF) shell-model to describe the initial nucleus, for which one obtains a ground state composed of single-particle orbitals for the nucleons.
Following the definitions of the lepton tensor $L_{\mu\nu}$ and recoil factor $f_{rec}$ in Ref.~\cite{Gonzalez-Jimenez:2021ohu} the differential cross section is 
\begin{align}
&\frac{\,d\sigma (E_m)}{\,d \lvert \vec{k}_f\rvert \,d\cos\theta_f \,d \lvert \vec{k}_N \rvert \,d\Omega_N} = \nonumber \\ 
&\frac{G^2_F\cos^2\theta_c}{2(2\pi)^4}   \left(\frac{\lvert \vec{k}_f\rvert^2 \lvert \vec{k}_N\rvert^2 M_B}{E_B f_{rec}}\right) L_{\mu\nu}\sum_{n,\kappa} \delta(E_m - E_{n, \kappa}) H_{n,\kappa}^{\mu\nu},
\label{eq:CS}
\end{align}
the hadron tensor $H^{\mu\nu}_{n,\kappa}$ is described in more detail in the next subsection. Here $n,\kappa$ denote the principal and relativistic angular momentum quantum number which uniquely label the states with single-particle energy $E_{n,\kappa}$. 
The knockout of a nucleon out of a certain shell leaves the residual system in an excited state with invariant mass determined by the single particle energy of the state through Eqs.~(\ref{eq:EM}-\ref{eq:PM}).

This shell model treatment is known to be a first approximation to more realistic missing-energy profiles. Experimental data shows that the discrete states obtain a width and are partly deoccupied with the missing strength appearing at larger missing energies~\cite{BENHAR:RevModPhyseep,RDWIA:Kelly, Dutta:PRCeepAu}. This is due to both long- and short-range correlations, and the effect of FSI.
A missing-energy profile that takes into account a background due to short-range correlations, based on the spectral function of Ref.~\cite{BENHAR1994, BENHAR2005}, was added to the same RDWIA approach used here in Ref.~\cite{Gonzalez-Jimenez:2021ohu}.
This is unnecessary for the present work, to isolate the effects of FSI we use the same shell model treatment also as input for the cascade model.
The pure shell model and spectral function approaches yield a similar shape for the hadron observables with the main effect of the spectral function being a reduction of the magnitude of the cross section as discussed in Ref.~\cite{VanOrden:2019}.

\subsection{Final-state potentials}
The hadron tensor for a nuclear shell is 
\begin{align}
    &H^{\mu\nu}_{\kappa,n}(Q,k_N) = \nonumber \\
    &\sum_{m_j,s_N} \left[J_{\kappa,n}^\mu(m_j,s_N,Q,k_N)\right]^\dagger J_{\kappa,n}^\nu(m_j,s_N,Q,k_N) 
\end{align}
where $m_j$ and $s_N$ are the angular momentum projection of the bound state and the spin of the final-state nucleon respectively. The total angular momentum of the bound state $j = \lvert \kappa \rvert - 1/2$. $Q$ and $k_N$ are four-momentum transfer $Q = k_i - k_f = \left(\omega,\vec{q}\right)$ and the outgoing nucleons four-momentum respectively. 
The hadron current is
\begin{align}
    &J_{\kappa,n}^\mu(m_j,s_N,Q,k_N) = \nonumber \\
    &\int\,d\vec{r} e^{i\vec{q}\cdot\vec{r}}~\overline{\Psi}(\vec{r},s_N,k_N) \mathcal{O}^\mu(Q) \psi_{\kappa}^{m_j}(\vec{r}).
    \label{eq:Jhad}
\end{align}
We will discuss in some detail the description of the final-state wavefunction $\Psi(\vec{r},s_N,k_N)$, we refer to Refs.~\cite{Gonzalez-Jimenez:2019ejf, Horowitz81} for a discussion of
 the bound state wavefunctions $\psi^{m_j}_\kappa$ obtained with the model of Ref.~\cite{NLSH, Horowitz81}, and the transition operator $\mathcal{O}^\mu(Q)$.
 For completeness we mention that we use the cc2 operator, with the form-factors of Ref.~\cite{Kelly04} for the vector current, and a dipole with cut-off mass $M_A=1.05~\mathrm{GeV}$ for the axial current where pion-pole dominance is used for the pseudoscalar form-factor.
 
 The final-state wavefunction with asymptotic momentum $\vec{k}_N$ in a central potential is obtained in a partial wave expansion
 \begin{align}
 &\Psi(\vec{r},s_N,k_N) = 4\pi \sqrt{\frac{E_N+M}{2E_N}} \sum_{\kappa, m_l, m_j} \nonumber \\
 &e^{-i\delta_\kappa^*} i^l \langle l~m_l,1/2~s_N~\lvert j~m_j \rangle Y_{l,m_l}^*\left(\Omega_N\right) \psi_\kappa^{m_j}\left(\vec{r},E_N\right)
 \end{align}
 where $Y_{l,m_l}(\Omega_N)$ are the spherical harmonics that describe the nucleon direction. The orbital angular momentum $l=\kappa$ if $\kappa > 0$ and $l = -(\kappa+1)$ if $\kappa < 0$, total angular momentum $j = \lvert\kappa\rvert - 1/2$.
 The solution of the central Dirac equation for a certain $\kappa$ has the form~\cite{Greiner}
 \begin{equation}\label{eq:radial}
\psi_{\kappa}^m(\vec{r}) = 
\begin{pmatrix}
g_\kappa(r)\phi^m_\kappa\left(\Omega_r\right) \\
if_\kappa(r)\phi^m_{-\kappa}\left(\Omega_r\right)
\end{pmatrix},
\end{equation}
where $r = \lvert\vec{r}\rvert$.
The angular dependence is described by spinor spherical harmonics
\begin{equation}
\phi_\kappa^m\left(\Omega_r\right) = \sum_{m_l,m_s} \cgc{l}{m_l}{1/2}{m_s}{j}{m_j} Y_{l,m_l} \left( \Omega_r \right) \chi^{m_s},
\end{equation}
with the two-component spinors
\begin{equation}
    \chi^{+1/2} =
    \begin{pmatrix}
1 \\
0
\end{pmatrix},~ 
\chi^{-1/2} =
    \begin{pmatrix}
0 \\
1
\end{pmatrix}.
\end{equation}

The radial wavefunctions $g_\kappa\left(r\right)$ and $f_\kappa\left(r\right)$ are solutions of the coupled Dirac equation with scalar (S) and vector (V) potentials
\begin{align}\label{eq:radDirac}
\frac{\,d g_\kappa}{\,d r} &= -\frac{\kappa}{r}g_\kappa + \left[E_N + S(r,E_N) - V(r,E_N) \right]f_\kappa \nonumber \\
\frac{\,d f_\kappa}{\,d r} &= +\frac{\kappa}{r}f_\kappa - \left[E_N - S(r,E_N) - V(r,E_N) \right]g_\kappa.
\end{align}
Both potentials include a strong interaction of finite range, and additionally the Coulomb potential is included in the vector potential such that the radial wave-functions behave like (phase-shifted) Dirac-Coulomb wave-functions at large $r$~\cite{Greiner}. The strong scattering phase-shift $\delta_\kappa$ and the normalization are determined by matching the solution of Eq.~(\ref{eq:radDirac}) to these asymptotic wave-functions at large $r$.

\begin{figure*}
\begin{minipage}{0.49\textwidth}
        \centering
            \includegraphics[width=0.98\textwidth]{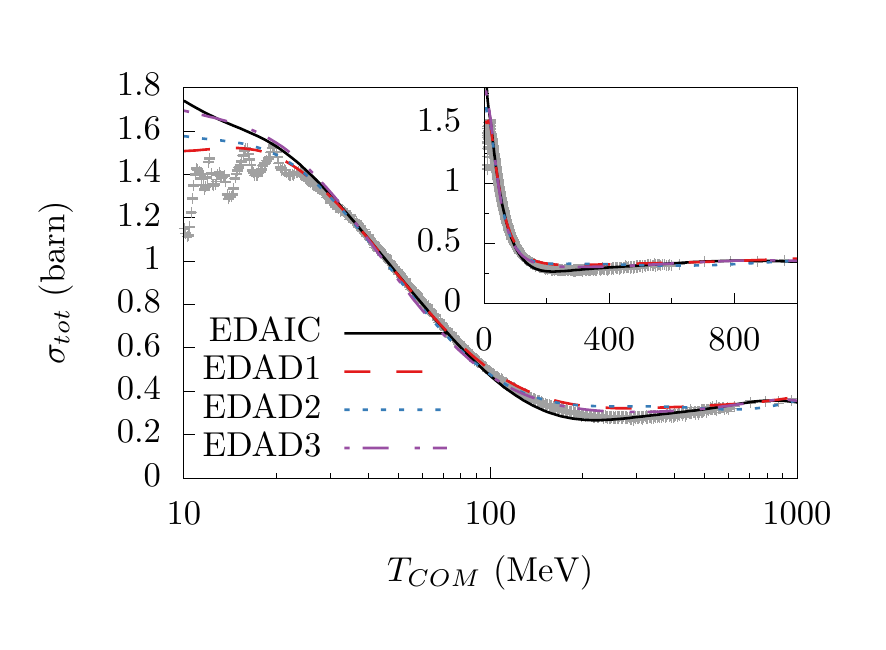} 

    \end{minipage}\hfill
    \begin{minipage}{0.49\textwidth}
      \centering
    \includegraphics[width=0.98\textwidth]{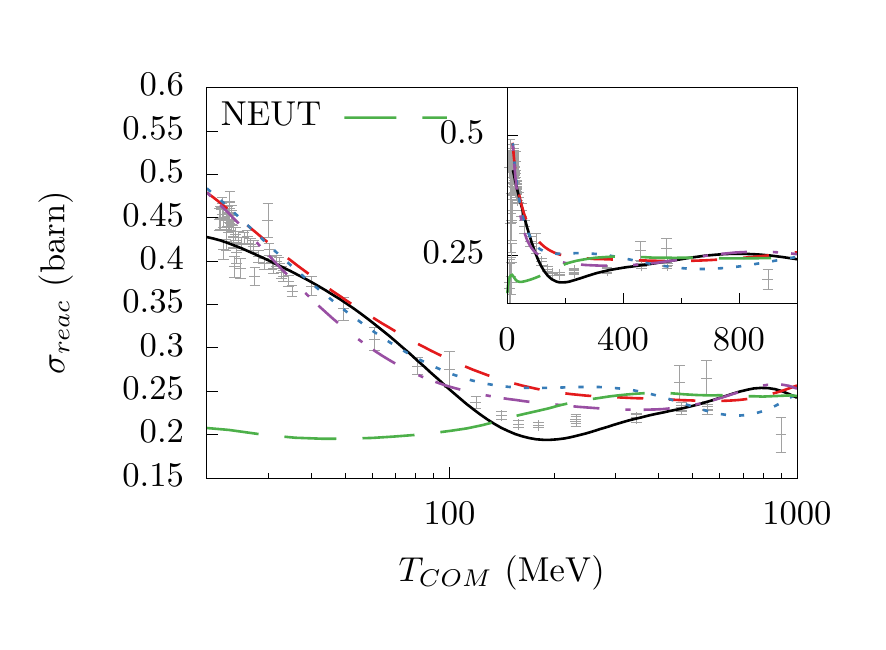}
    \end{minipage} 
    \caption{Total cross section for neutron scattering (left) and reaction cross section for proton scattering (right) off carbon. Experimental data from Refs.~\cite{ReacpC1,ReacpC2,ReacpC3,ReacpC4,totnC1,totnC2,totnC3} are compared to the results obtained with the different optical potentials from Ref.~\cite{Cooper:1993} and the NEUT result. The reaction cross section predicted by NEUT is obtained from Ref.~\cite{NEUT:2021EPJST}. The insets show the same results on a linear scale. The data were obtained from the EXFOR database~\cite{OTUKA:EXFOR}. }
    \label{fig:N_C12_sc}
\end{figure*}
The energy-dependent scalar and vector optical potentials used in this work are obtained from the analysis of proton-nucleus elastic scattering.
We use the energy-dependent A-independent (EDAI) fits for ${}^{12}$C, ${}^{16}$O, and ${}^{40}$Ca of Ref.~\cite{Cooper:1993}. The fits include scattering data with proton kinetics energies up to $1040~\mathrm{MeV}$ and down to $29$, $23$, and $21$ MeV for carbon, oxygen, and calcium respectively.
In order to cover the whole phase space we extrapolate the potentials also to lower values, however it should be understood that the potentials are not constrained in that range.
Moreover the assumption underlying the optical model, namely a dense continuum of inelastic channels, is expected to break down for small energies where inelastic interactions that proceed through discrete energy-levels become important.

The imaginary part of the optical potential absorbs flux that is lost to inelastic interactions and which is not part of the signal in elastic $p-A$ scattering. The optical theorem relates the total cross section (elastic plus inelastic) to the imaginary part of the elastic amplitude in the forward scattering limit, and such potentials reproduce the total cross section well~\cite{Cooper:1993,Cooper:2009}.
We show the results for the total cross section for neutron scattering off ${}^{12}$C with different potentials in the left panel of Fig.~\ref{fig:N_C12_sc}, comparisons to total cross sections off other nuclei can be found in Ref.~\cite{Cooper:1993}. In addition to the EDAIC potential that is fit exclusively to scattering off carbon which we will use in this work, we also include the $A$-dependent (EDAD) fits which are constrained by scattering off several nuclear targets.
The different potentials yield similar results for kinetic energies larger than approximately $30~\mathrm{MeV}$. The cross sections obtained with different fits diverge from each other for smaller energies outside of the fitted range.
In the NEUT cascade model, as described in Refs~\cite{NEUT:2021EPJST} and explained below, elastic scattering off the whole nucleus is not modeled and as such the total cross section is also not available.

In NEUT, the total reaction cross section is modeled by considering elastic and inelastic interactions with the constituent nucleons in the nucleus.  
We compare the reaction cross section for $p-{}^{12}\mathrm{C}$ scattering in NEUT, obtained from Ref.~\cite{NEUT:2021EPJST}, with ROP calculations and data in the right panel of Fig.~\ref{fig:N_C12_sc}.
The difference between the cross sections obtained with different potentials is larger in this case. Still the data is reproduced reasonably by the different approaches down to small kinetic energy. It is seen that the reaction cross section of NEUT is comparable to ROP and the data for kinetic energies larger than approximately $100~\mathrm{MeV}$, but does not reproduce the sharp increase of the cross section at smaller energies. Different cascade models however do reproduce the low-energy peak of the reaction cross section~\cite{ValidationmethodsINC:2021,Isaacson:cascade2021} 

Modeling FSI with the complex ROP is well suited for a description of {\it exclusive} one-nucleon processes where the missing energy $E_m$ is restricted to a narrow range, and only the nucleons which do not suffer inelastic interactions define (the bulk of) the experimental signal.
To compute the inclusive cross section, one should instead retain the inelastic interactions. The total inclusive cross section can be extracted in principle consistently from the ROP using the (relativistic) Green's function technique~\cite{BOFFI1993,Giusti2012MB}.
In Refs.~\cite{Gonzalez-Jimenez:2019ejf, Gonzalez-Jimenez19,Kim03, Butkevich:eeprime2020} and others, a simpler approach is used in which the inclusive cross section is described by using only the real part of the optical potential, which is referred to as the rROP.
This approach is found to yield realistic results for the inclusive cross section up to large momentum transfer. We will use the latter approach in this work.

\section{Hadron variables in T2K \label{sec:T2KTP}}
We look for events in which a single proton with four-vector $k_p$ is detected in the final-state in coincidence with a muon $k_\mu$.
We have generated events for this signal with the T2K flux~\cite{T2K:flux2016} according to RDWIA calculations with different potentials.
The events are characterized by four-vectors $k_\nu$, $k_l$, $k_p$ which are distributed according to the differential cross section of Eq.~(\ref{eq:CS}) weighted with the normalized T2K flux.
We consider following models:
The relativistic plane wave impulse approximation (RPWIA), in which FSI are neglected completely by treating the outgoing particle as a relativistic plane wave; The rROP, discussed above, uses the real part of the optical potential fitted to $p-A$ scattering data; And the ROP calculation that uses the full optical potential, i.e. including the imaginary part.

We have introduced the events obtained with the rROP and RPWIA in the NEUT cascade model. In this cascade approach, the single proton that is present in the event used as input is the only particle that is affected, i.e. the lepton kinematics are unchanged.
Although one can not formally disentangle the elastic propagation of the nucleon from the inelastic contributions as simply the imaginary and real part of an empirical optical potential, it can clearly be argued that supplying the cascade with the rROP events does not pose a double counting issue. 
The rROP potential modifies the dispersion relation of the outgoing nucleon in the nuclear interior. In this propagation, no other nucleons are explicitly knocked out nor any additional particles are created, i.e. the nucleon exchanges momentum, but not energy, with the residual system.
In the present cascade model no such effects are included, and every interaction exchanges energy and momentum with the constituents of the nucleus.

One may further motivate the use of the rROP approach as input for the cascade model from a more practical point of view.
As mentioned above, the cascade model does not affect the inclusive cross section, but only the composition of the hadronic final state.
This is at variance with the RDWIA calculations, where a different final-state potential has an effect on the lepton variables. Among other things, the real potential introduces a $q$ and $\omega$ dependent shift of the quasielastic peak compared to RPWIA calculations, in agreement with electron scattering data as shown e.g. in Refs.~\cite{Gonzalez-Jimenez:2019ejf,Gonzalez-Jimenez19}.
It is important to feed the cascade model with a calculation which gives good results for the inclusive cross section such as the rROP. Indeed, the cascade model redistributes strength into specific open channels in the final-state and upon integration over all channels one wants to recover the correct inclusive result.

\subsection{Selecting elastic events}\label{sec:selection}
\begin{figure*}
\includegraphics[width=0.485\textwidth]{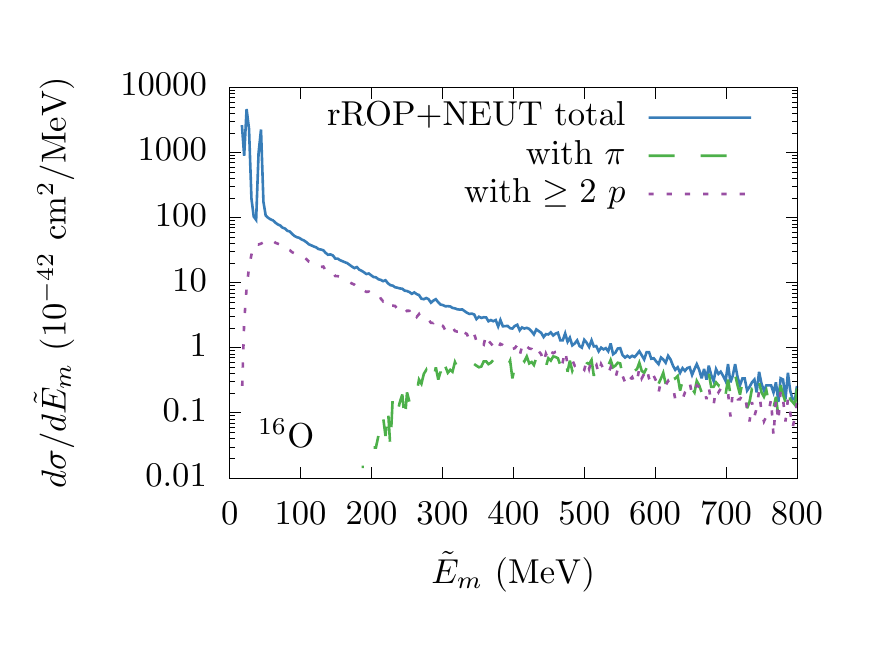}
\includegraphics[width=0.485\textwidth]{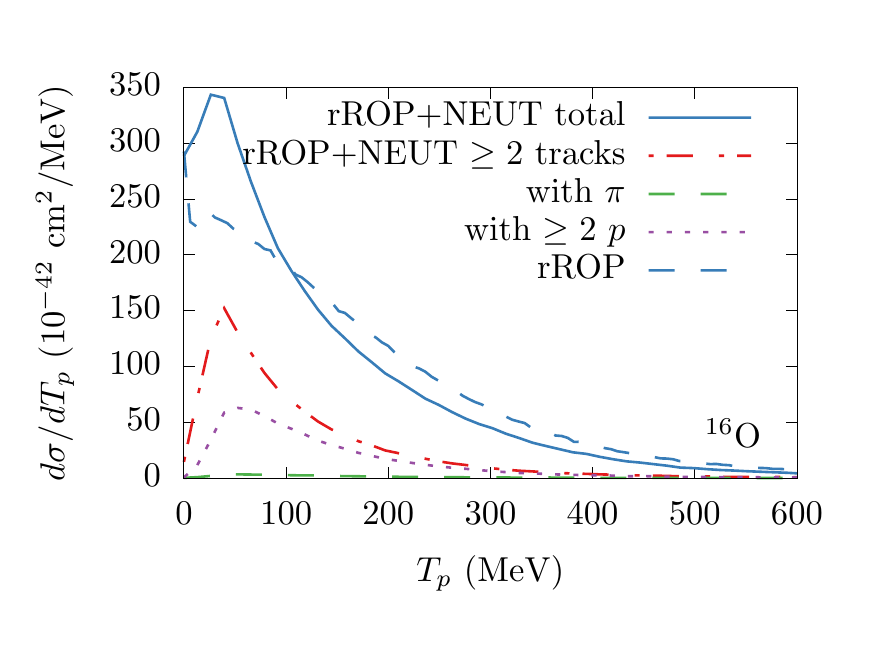}
\caption{The composition of the final-state from the cascade over a large missing energy region in scattering off oxygen, with the incoming energies distributed according to the T2K ND280 $\nu_\mu$ flux~\cite{T2K:flux2016}. We show the events in which more than 1 hadron track is present, and the subset in which 2 or more protons make it out of the nucleus, or in which a pion is involved (which may or may not make it out).
}
\label{fig:NEUTEm}
\end{figure*}

The ROP calculation can serve as a benchmark for the hadron kinematics, it should give the correct reduction of the cross section in a quantum-mechanical manner, however it does not tell us 'where the particle goes' after undergoing a secondary interaction, instead the flux lost to such interactions is removed.

In order to make a meaningful comparison of ROP cross sections with the events resulting from the cascade model we need to make a selection of a class of events.
Firstly to select events and the kinematics from the cascade which correspond to a $1p1\mu+X$-signal we select for every event the most energetic proton that makes it out of the nucleus.

We then make a selection on the events from the cascade model which have not been affected by inelastic FSI. We propose two methods, the first based on the classification of the events in the NEUT cascade, and the second based solely on the knowledge of the kinematics of an event.

For the first we use the NEUT output which yields for every event a number of hadron 'tracks' that follow interaction points in the nucleus. 
When only one hadron track is present in an event, the original proton leaves the nucleus without any interactions. When multiple tracks are present FSI has occured and the original proton will generally change energy and direction. In this cascade model, the outgoing proton can only exchange energy-momentum with the constituents of the nucleus, and not the nucleus as a whole, hence in these cases other hadrons will explicitly be present in the final-state.
This means that a selection on these single track events should correspond to an 'elastic' signal, which in this cascade model is simply the case where nothing happens.

\begin{figure*}
\includegraphics[width=0.485\textwidth]{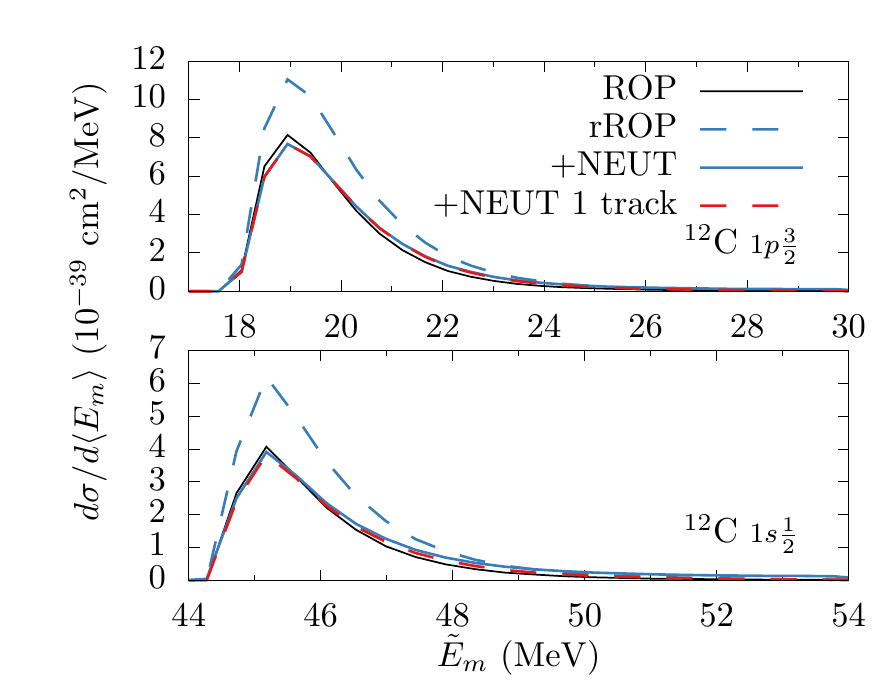}
\includegraphics[width=0.485\textwidth]{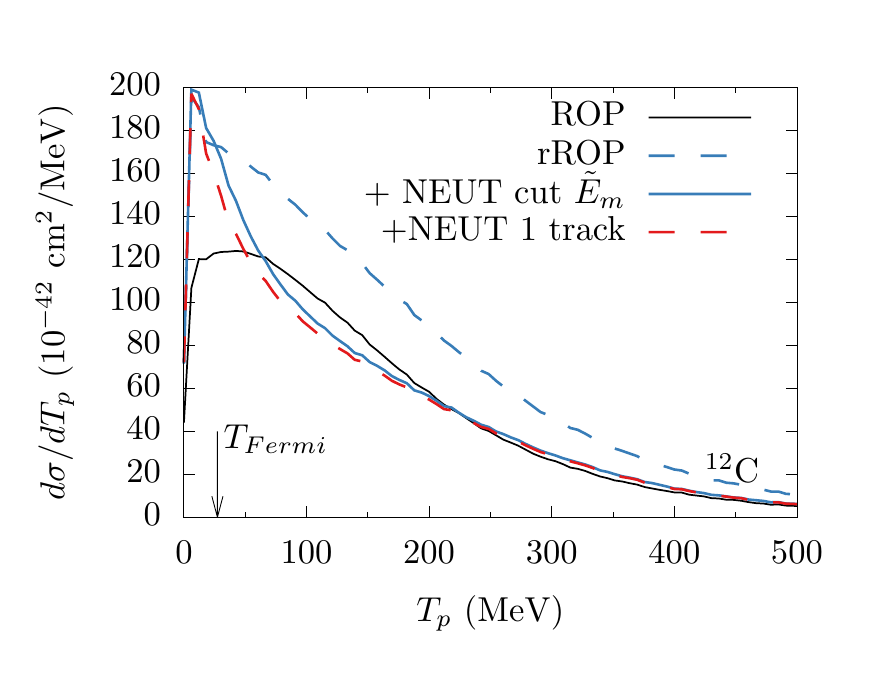} 
\caption{Comparison of the original rROP input (dashed lines) with the result after the NEUT cascade is applied  (solid blue) and the ROP calculation (black). In the NEUT results shown in the right panel only events with $\tilde{E}_m$ corresponding to the region shown in the left panels are retained. In the results shown by red dashed lines no cut in missing energy is made, but instead the selection of '1-track' events in NEUT is made.}
\label{fig:NEUTshells}
\end{figure*}

On the other hand, one may want to make a selection of events based purely on kinematics instead of the classification used in a model.
From the viewpoint of a neutrino experiment, the variables that we can look at are only the nucleon and lepton kinematics.
However, as we are simulating we have the information of the true incoming energy, and we can thus define for every event a missing energy as
\begin{equation}
\label{eq:Emiss}
\tilde{E}_{m} = E_\nu - E_\mu - T_p.
\end{equation}
While the missing energy defined in this way is not directly measurable in a $1p1\mu$ event unless the incoming energy is known, it does correspond to an energy which can in principle be characterized by  additional knowledge of the content of the final-state, e.g. if a pion is produced one has at least $m_\pi$ in missing energy.
The tilde denotes that this definition of missing energy does not take into account explicitly the kinetic energy carried away by the residual hadronic system as in Eq.~(\ref{eq:EM}).

In Fig.~\ref{fig:NEUTEm}, we show the distribution of events when the NEUT cascade is fed with the rROP model in terms of missing energy $\tilde{E}_m$ and the leading proton momentum.
We show specifically the events in which a pion is found in the cascade (which may or may not make it out of the nucleus), which are completely negligible. 
We find that the events with more than 1 track tend to be concentrated mostly at low values of $T_p$. This is because the originally higher energy proton loses energy in a collision. For about half of these events an additional proton is predicted to be present in the final state.

Because we are using a shell model the RDWIA events used as input are all concentrated in narrow peaks in $\tilde{E}_m$.
We make a selection of events after applying the NEUT cascade, in which only those events that correspond to these regions of $\tilde{E}_{m}$ are retained.
The idea is that inelastic interactions would change the energy of the nucleon more significantly, in which case the event ends up in a different missing-energy region.

In the left panels of Fig.~\ref{fig:NEUTshells} we show the cross section in terms of $\tilde{E}_m$ in the shell model region. 
We see indeed that the selection of events with only 1 hadron track is practically identical to the full NEUT result in the region below the shell model peaks. The agreement is perfect for the $p$-shell, this is because interactions in the cascade will generally increase $\tilde{E}_m$. In the higher $\tilde{E}_m$ $s$-shell region the additional strength in the full NEUT result compared to the 1-track selection is very small.
The right panel of Fig.~\ref{fig:NEUTshells} compares the cross section in terms of proton kinetic energy obtained with the 1-track selection to the result in which cuts are made such that only the $\tilde{E}_m$ regions shown in the left hand panels are included.  
We see indeed that this procedure reduces the NEUT result practically to the 1-track result with a minimal contribution of multi-track events.

\subsection{Comparison of the NEUT cascade and ROP}\label{sec:comparisonC}
With the event selection explained in the previous section the inelastic contribution is removed from the rROP+NEUT results, which should hence be comparable to the ROP calculations.
The comparison is made for the proton kinetic energy $T_p$ in the right panel of Fig.~\ref{fig:NEUTshells}, the rROP+NEUT results match the ROP results for high energy protons with this event selection.
Both the ROP and the cascade model are constrained by p-C scattering data, and as seen in Fig.~\ref{fig:N_C12_sc} both give a similar magnitude for the reaction cross section in this kinetic energy region. It should however be appreciated that both models include these constraints in a very different manner, and that their agreement is non-trivial.
For smaller kinetic energies the NEUT results are significantly larger than the ROP cross section. This can likely be understood from the results for the reaction cross section as well, in NEUT the sharp rise of the cross section at small energies, which is apparent in the ROP results, is not present.
Other cascade models do reproduce the rise of the reaction cross section at low energies with varying degrees of accuracy~\cite{ValidationmethodsINC:2021,Isaacson:cascade2021}, hence this behaviour is not expected to be universal in event generators. In particular, a similar study as the one performed here with the NuWro event generator~\cite{NuWro:FSI} gives a stronger decrease of the cross section at small $T_p$, yielding results that are more comparable to the ROP~\cite{Niewczas:prep}.
Additionally it is notable that for energies smaller than approximately the kinetic energy corresponding to the Fermi energy (indicated by the arrow in Fig.~\ref{fig:NEUTshells}), the rROP+NEUT 1-track cross section is identical to the rROP result used as input.
As explained in Ref.~\cite{NEUT:2021EPJST}, in the NEUT cascade a nucleon's momentum should be larger than the (local) Fermi momentum after interaction. This implies that the lowest energy nucleons do not interact and leave the nucleus unimpeded.

While, as we have discussed in Section~\ref{sec:formalism}, one should proceed with caution when extending the ROP model to small nucleon energies, it is reasonable to assume that this description is more robust than the cascade for small energies. For one, the approach is quantum mechanical which is important at low kinetic energies where the nucleon's reduced wavelength becomes comparable to the size of the nucleus. In this regime collective degrees of freedom and absorption become relevant, these are effectively captured in the empirical optical potentials, but are not present in the cascade in which only scattering with constituents is considered.
Additionally, although total cross sections obtained with different potentials deviate at small energies as shown in Fig.~\ref{fig:N_C12_sc}, the nucleon-nucleus cross sections are described more accurately by the ROPs in the low-energy region than by NEUT. 
In the following sections we examine the agreement and differences between the cascade and ROP approaches in some detail. We pay attention to the dependence on the position at which a nucleon is introduced in the cascade, the scaling towards heavier nuclei, and the model that is used to supply the cascade with events.

\subsection{Nuclear density \label{sec:rdep}}
\begin{figure*}[t]
\includegraphics[width=\textwidth]{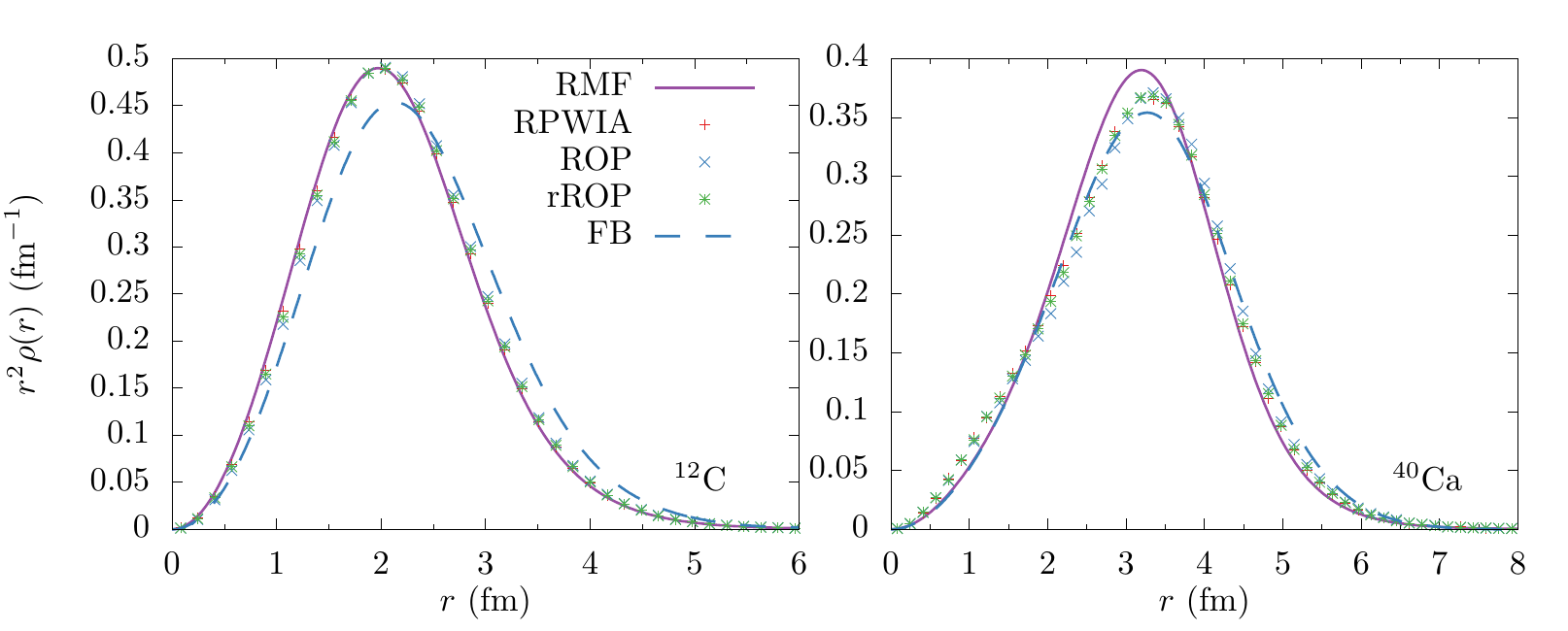}
\caption{Distribution of the radius at which events are introduced in the NEUT cascade in accordance with calculations with different final-state potentials are shown by crosses. The radial distributions obtained from the (point-like) neutron-density in RMF (solid lines), and from the experimental charge-density parametrized by Fourier-Bessel (FB) coefficients (dashed lines) are also shown~\cite{DEVRIES1987}. All results are normalized such that $\int \,d r r^2\rho(r) = 1$.
The left panel shows the results for carbon while the right shows the result for calcium.}
\label{fig:RNEUT}
\end{figure*}

We briefly examine the dependence of the NEUT results on the position at which the nucleon is introduced in the cascade model.
The position at which the nucleon should enter the cascade is not easy to answer from the RDWIA approach.
In these calculations the overlap of initial and final-state wave functions is calculated over the whole space. 
In any case it is natural that the position at which events are introduced in the cascade should be proportional to the nuclear density. We will here compare the results for the cross section when the events are introduced according to a consistent, realistic nuclear density profile with those obtained for a simple uniform density. 
More elaborate approaches could be considered in future work, including for example dependence on $T_p$ or missing momentum, which may be important especially in the low-$T_p$ region.  

\begin{figure}
\includegraphics[width=\columnwidth]{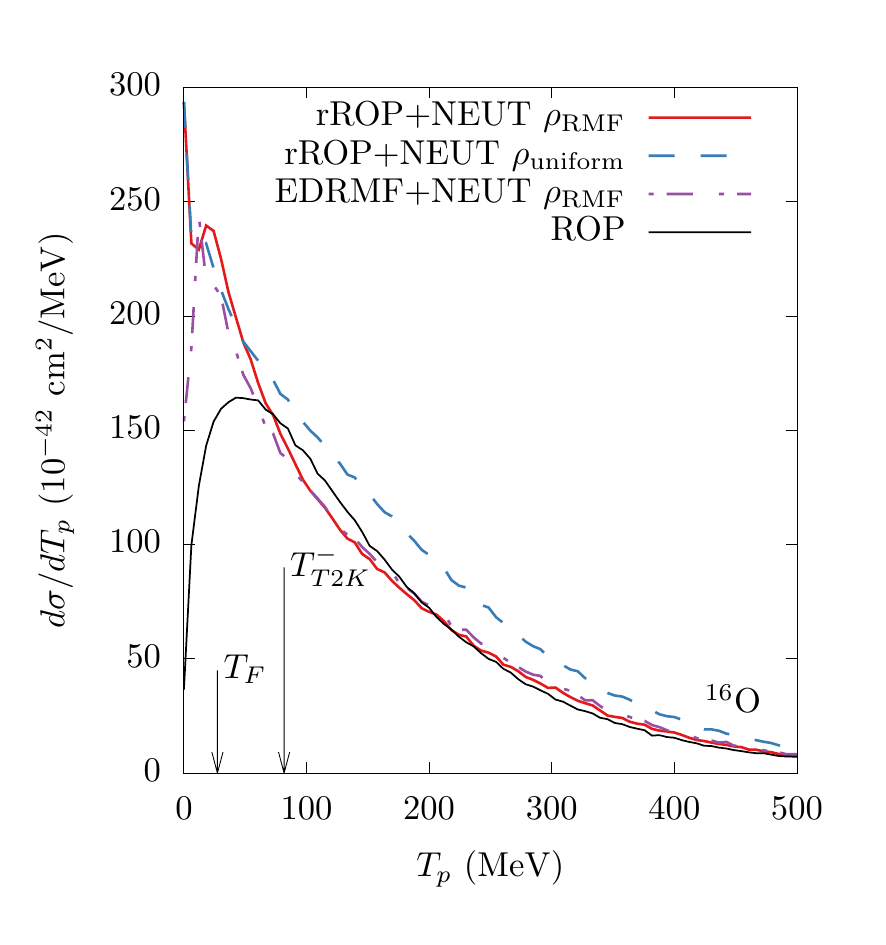}
\caption{Cross section averaged over the T2K flux, differential in the leading protons kinetic energy. ROP calculations are compared to NEUT results with a cut in missing energy to isolate elastic events. The arrows show the kinetic energies corresponding to the Fermi energy ($T_F$) and the proton detection threshold in T2K ($T^{-}_{T2K}$). }
\label{fig:Tp_rho_O16}
\end{figure}

For consistency we will make use of the nuclear density obtained within the RMF, which yields reasonable results, especially for even-even nuclei with large mass~\cite{NLSH, Horowitz81}.
We have used an approach that makes full use of the shell-model description. The radius at which a nucleon is introduced in the cascade is distributed according to the density of the corresponding shell from which it originates.
In general, it is not necessary that the sum of the density corresponding to the shells weighted with the cross section is the same as the ground-state density.
For example if $\omega$ is smaller than the binding energy of a shell, the cross section will not be sensitive to the density of that shell in this approach.
Fig.~\ref{fig:RNEUT} shows the distribution of radii at which the event is introduced in this approach, and compares it to the distribution obtained from the ground-state neutron density in RMF. 
One sees that the distributions obtained with different models for the final-state wave functions are practically the same. They differ only minimally from the RMF point-neutron density for oxygen, and slightly more for calcium.
This means that, for flux-averaged results that are integrated over the whole phase space, the different shells contribute to the total cross section with a weight that is proportional to their relative occupancy. 
Because of this, and as the overall differences between models are negligible, it seems reasonable that one could instead use the overall neutron density, either from RMF or some other realistic model, instead of a different one for every shell.
For completeness, we also show the distribution obtained from the experimental charge density from Refs.~\cite{DEVRIES1987, Chdenssite}.

The results using this approach, including again a cut on $\tilde{E}_m$ as introduced in section~\ref{sec:selection} to select elastic events, are shown in Fig.~\ref{fig:Tp_rho_O16}.
The results are compared to those obtained when the events are introduced in the cascade according to a sphere with uniform density and radius $5~\mathrm{fm}$.
We find a good agreement of the ROP and rROP+NEUT results when the RMF densities are used, but the reduction of the cross section is too small for the uniform density.
The reduction due to FSI is increased compared to the uniform distribution, because more nucleons are introduced deeper inside the nucleus and thus have a larger chance for interaction.

The ROP and NEUT+rROP results including the cut on missing energy agree well for kinetic energies above approximately $100~\mathrm{MeV}$, the agreement is qualitatively similar for oxygen as for carbon.
An arrow $T^{-}_{T2K}$ is added to Fig.~\ref{fig:Tp_rho_O16}, corresponding to a proton with a momentum of $450~\mathrm{MeV}$, which is the lower bound for the T2K analysis.

For low nucleon energies quantum-mechanical effects become important, and hence the ROP should be the natural method to describe FSI as discussed previously.
However, a lack of Pauli blocking and spurious contributions to the matrix element, both due to the fact that initial and final states are not described consistently, can affect the cross section in this region~\cite{Boffi:1982id, Nikolakopoulos19,Atti83, Johansson00, Jachowicz:JPG2019}. 
For this reason we include in Fig.~\ref{fig:Tp_rho_O16} also results obtained by feeding the cascade with events generated with the Energy-Dependent RMF (ED-RMF) cross section. 
The empirical ED-RMF potential gives similar results as the rROP when the nucleon energy is large, but reduces to the same RMF potential used to describe the initial state when the nucleon has small energy~\cite{Gonzalez-Jimenez19}.
In this way spurious non-orthogonal contributions to the matrix element are not present for low energies, which is where their effect is largest~\cite{Gonzalez-Jimenez:2019ejf,Gonzalez-Jimenez19}.
One sees that the ED-RMF and rROP results are very similar, with the ED-RMF yielding a slightly smaller cross section for small $T_p$.

\subsection{$A$-dependence\label{sec:Adep}}
\begin{figure}
\includegraphics[width=\columnwidth]{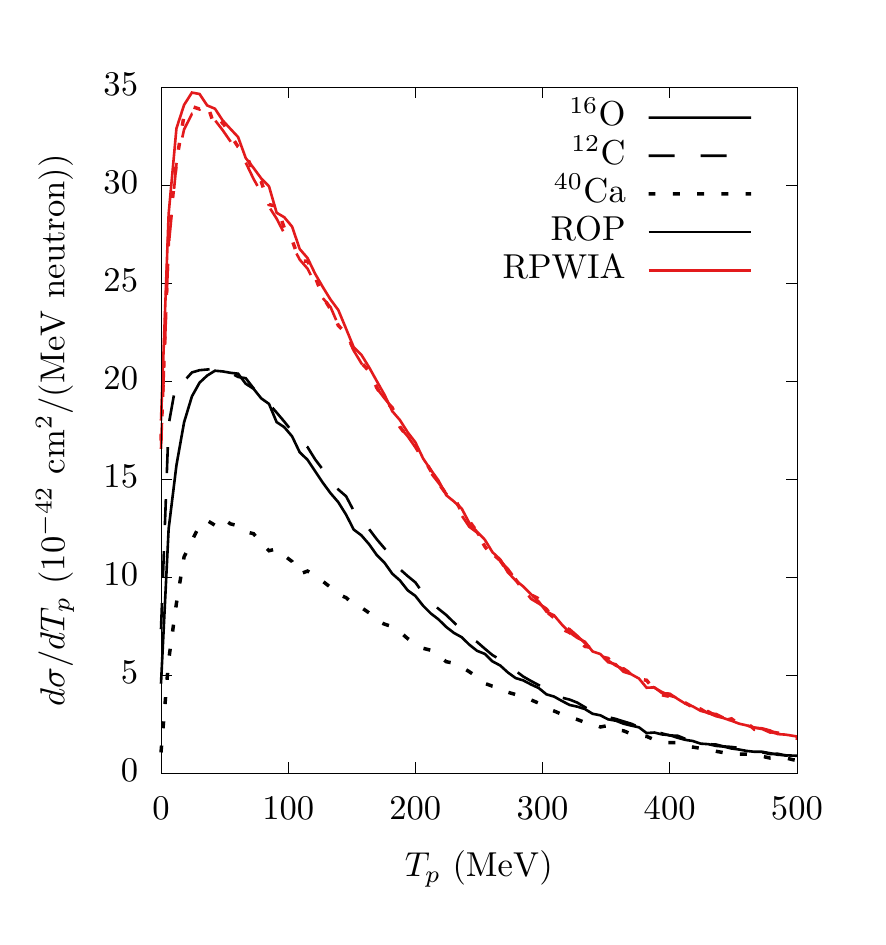}
\caption{Comparison of ROP (black) and RPWIA (red) results for the T2K-flux averaged cross sections for different nuclei.}
\label{fig:AdepRPWIA}
\end{figure}

To check if the agreement found for oxygen and carbon is a result of fine-tuning of the cascade to carbon data, or rather a more robust result, we extend the comparison with cross sections for calcium.
Additionally we include results where the input to the NEUT cascade are RPWIA cross sections.
First, in Fig.~\ref{fig:AdepRPWIA}, we show the RPWIA and ROP results normalized per target neutron.
The RPWIA results are practically identical for all nuclei, but this naive scaling disappears when the optical potential is included.
One finds that the reduction of the cross section compared to the RPWIA result is larger for calcium than for oxygen and carbon.
This should be expected, from electron scattering measurements it is well known that the nuclear transparency decreases with mass ~\cite{Dutta:PRCeepAu,Rohe:2005}.

\begin{figure*}
\includegraphics[width=\textwidth]{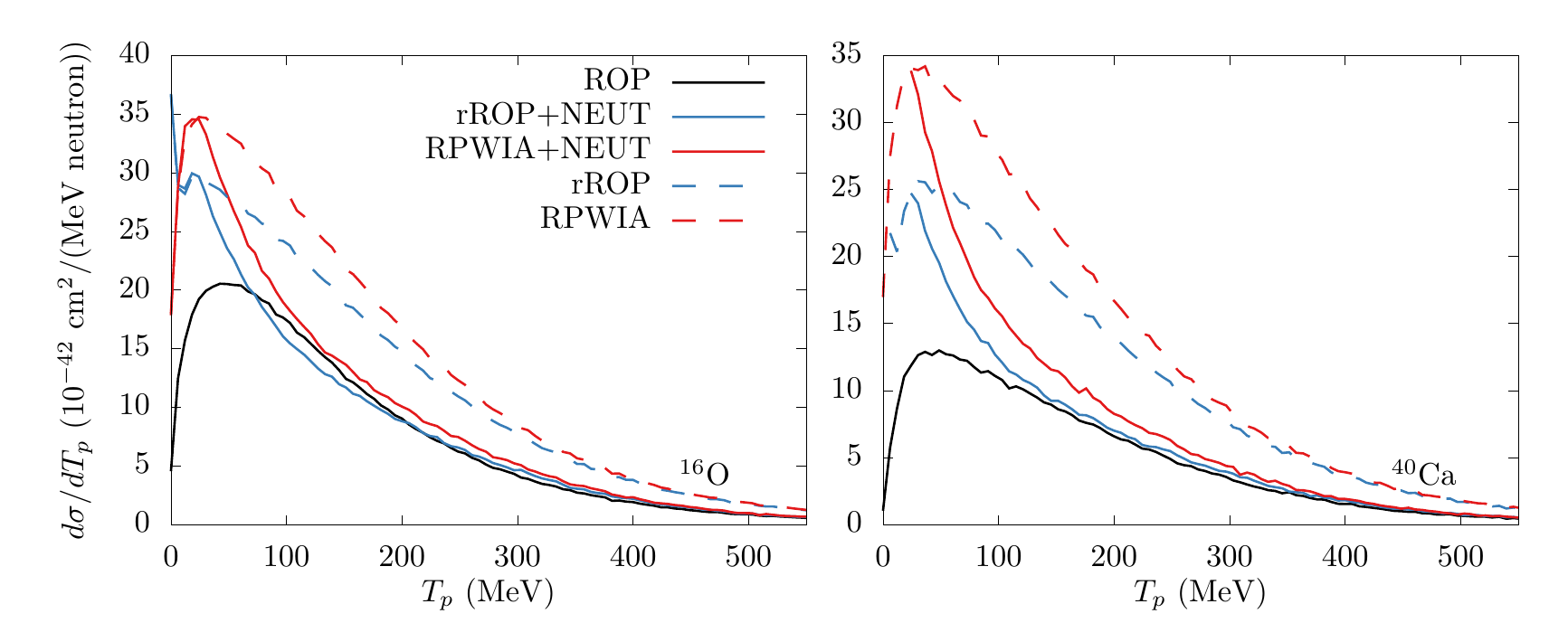}
\caption{Cross section in terms of the leading protons kinetic energy averaged over the T2K flux. All results include a cut in missing energy to isolate elastic events. ROP results are compared to the NEUT results when using rROP or RPWIA as input to the cascade. The results of the models before application of the cascade are shown by dashed lines.}
\label{fig:NEUTAdep}
\end{figure*}
In Fig.~\ref{fig:NEUTAdep} we show the result when the NEUT cascade is applied. As we find that oxygen and carbon give very similar results, we only show the oxygen and calcium cross sections.
Again, in the NEUT results a cut in $\tilde{E}_m$ is included to remove the inelastic contributions such that the ROP and NEUT results are comparable.
We find that the agreement between ROP and rROP+NEUT is quantitatively similar in calcium to the results for oxygen. 
While the rROP+NEUT results come very close to the ROP for ${}^{40}$Ca when $T_p > 100~\mathrm{MeV}$, this is not so much the case for the RPWIA.

While the agreement for 3 nuclei of course does not make for a significant set to determine the $A$-dependence, a disagreement would show that either additional degrees of freedom apart from the nuclear density should play a significant role in the cascade model, or that the cascade model might be tuned to only reproduce results for the lighter oxygen and carbon nuclei.
The results found here support that, when looking at the hadron variables for sufficiently large kinetic energies, a reasonable value for the nucleon-nucleon cross section with a realistic density dependence are sufficient to reproduce the nuclear transparency found in the ROP. Granted however that the cascade model is fed with the rROP RDWIA results to begin with.
These are also the essential degrees of freedom in the Relativistic Multiple Scattering Glauber Approximation (RMSGA) approach of Refs.~\cite{Ryckebusch:2003fc}, which was compared to the RDWIA in Ref.~\cite{Martinez06}, and yields similar results  for sufficiently large values of $T_p$.

\section{Combining lepton and hadron information \label{sec:T2Kdata}}
\begin{figure}
    \includegraphics[width=\columnwidth]{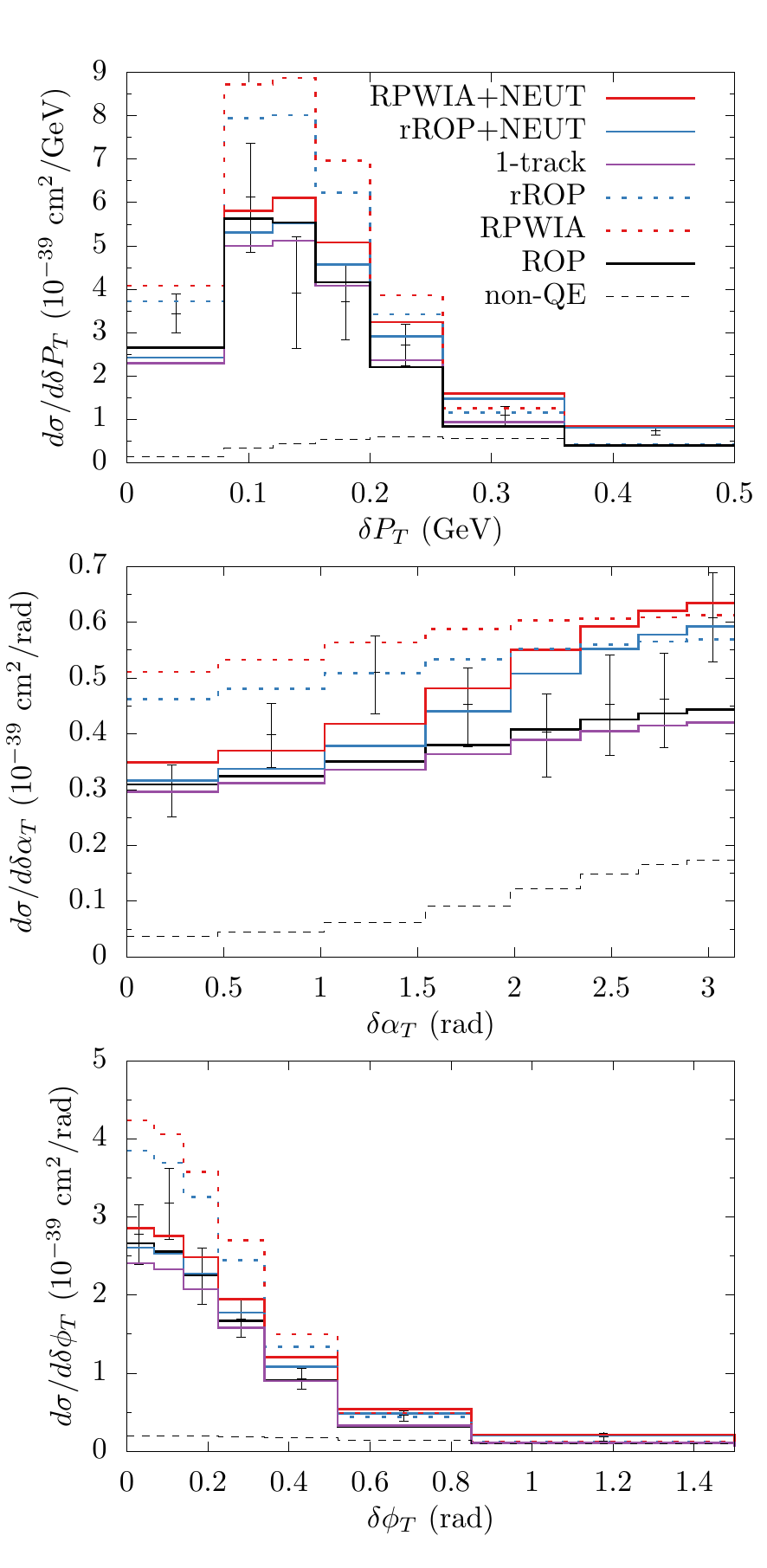}
    \caption{Results for the single transverse variables compared to the T2K data of Ref.~\cite{T2K:2018rnz} where the non-QE results of Ref.~\cite{Bourguille:2020bvw} (dashed lines) have been added to all calculations.}
    \label{fig:T2K_STV}
\end{figure}
In the previous section we focused on the energy of the outgoing proton, integrating over the lepton kinematics.
We found that for high $T_p$, the rROP+NEUT result resembles the ROP result very closely.
In the cascade model approach the FSI affects the the outgoing hadron kinematics only, and there is no dependence on the energy/momentum transferred to the nucleus.
However, the hadron current of Eq.~\ref{eq:Jhad}, and hence the description of FSI in the RDWIA approach, depends not only on the energy of the outgoing nucleon but also on $\vec{q}\cdot\vec{k}_N$, i.e. the magnitude and direction of the momentum transfer with respect to $\vec{k}_N$.
These differences cannot be readily discerned by considering (flux-folded) single-differential cross sections in terms of $T_p$.

Observables that combine lepton and hadron information are e.g. the transverse kinematic variables described in Refs.~\cite{Dolan:2018zye}.
In Fig.~\ref{fig:T2K_STV} we compare cross sections in terms of these variables to T2K data~\cite{T2K:2018rnz}. 
The experimental signal is defined as an event with no pions where one muon and at least one proton are detected in coincidence, such that other interaction mechanisms than single-nucleon knockout contribute to the data. For a full comparison, we have included the contributions of additional interaction mechanisms to the experimental signal using the results of Ref.~\cite{Bourguille:2020bvw}. The only change in the different calculations shown is the one-nucleon knockout contribution.
The additional interaction strength is shown separately in Fig.~\ref{fig:T2K_STV}, it stems mainly from 2-particle 2-hole excitations evaluated with the model of Ref.~\cite{NIEVES2011}, and from resonance excitation which does not lead to a detectable pion~\cite{NEUT:2021EPJST, ReinSeghal,Rein07}.

The calculations include the kinematic cuts implemented in the T2K data~\cite{T2K:2018rnz}. This implies in particular that the proton momentum is larger than 450 MeV and smaller than 1 GeV, i.e. in the region where the rROP+NEUT and ROP models give similar results for the $T_p$ distribution if the missing energy is restricted to the same region.
We do not apply the cut in missing energy for the cascade model results in this case. We do show the rROP+NEUT results restricted to only events with 1 track separately. The shape and magnitude of the cross sections in that case are similar to the ROP result.

The large errorbars make it difficult to asses the quality of the comparison to data, but some noteworthy trends emerge in the comparison of the different approaches.
The rROP and RPWIA results are clearly very large in the region of small $\delta P_T$ and $\delta\phi_T$. 
The cascade model redistributes these events and many of them end up below the threshold for proton momentum, which leads to a reduction of the cross section. A rather significant number of events remain in the experimental phase space, and these appear at high $\delta P_T$. 
From the comparison to $\delta \alpha_T$ one sees a significant shape difference between the microscopic calculations and the 1-track results on the one hand, and the results with the full cascade on the other hand.
The rise of the cross section with increasing $\alpha_T$ is more significant in the full cascade model results. The increase with $\delta \alpha_T$ in the other results stems from the addition of the non-QE cross section.
This shape is supported by the data, but the large error bars do not allow to draw any definite conclusions. A similar dependence on $\delta \alpha_T$ has been found in the MINERvA experiment~\cite{MINERvA:2018hba} and is well described by the NEUT cascade as shown in Ref.~\cite{Bourguille:2020bvw}. 

The RPWIA and rROP results tend to be similar in shape, whether or not the cascade model is used, and seem to differ mostly in terms of normalization for $\delta \alpha_T$.
For $\delta P_{T}$ and $\delta \phi_T$, the differences in magnitude between RPWIA and rROP are concentrated in low values of the variables, while they seem to converge somewhat for high values.
The difference between the RPWIA and rROP models is relatively small because the kinematic region that is probed does not include nucleon momenta smaller than 450 MeV, where these approaches deviate more significantly.

\section{Electron scattering data \label{sec:electron}}
The discrepancy between the rROP+NEUT and ROP results for an elastic signal is largest at small $T_p$. Differences in both the shape and magnitude of the $T_p$ distribution are found for $T_p < 100~\mathrm{MeV}$.

Confronting different models to neutrino data, for example in terms of the single-transverse variables presented in section \ref{sec:T2Kdata}, may allow to discriminate between different models, but such comparisons are not necessarily well suited to isolate the effects of FSI.
This is partly due to limited statistics in neutrino data, but mostly because neutrino beams span a broad energy range such that different interaction mechanisms cannot be easily separated.
Electron scattering data should be able to provide more stringent constraints and insights in this respect.
When the incoming energy is known, the missing energy can be restricted to eliminate interaction channels beyond quasielastic scattering in order to probe the effects of FSI in a controlled manner.

Measurements of the $(e,e'p)$ process on a variety of nuclei have been performed, which may be used to inform the treatment of FSI.
The most direct measurements of FSI in electron scattering come in the form of nuclear transparencies~\cite{PhysRevC.45.780,Rohe:2005,Garrow:2001di,PhysRevLett.80.5072,ONEILL1995}. 
The reported transparency is defined as the ratio of the number of protons measured experimentally to a theoretical expectation which does not include FSI.
The dependence on the phase space and the specific kinematic setup is expected to largely cancel in the ratio, and for large values of $T_p$ the measured transparencies are indeed found to be approximately constant.
It is important to keep in mind that the reported transparency is computed with respect to the expectation of a model.
The models used in these analyses are often based on the factorized PWIA, with a spectral function that is constrained to the measured data.
The spectral function may include the effect of short-range correlations (SRC), and in particular the resulting depletion of strength from single-particle orbitals.
As emphasized in Ref.~\cite{Rohe:2005}, the number of nucleons missing due to FSI cannot be distinguished from the depletion of the single-particle orbits due to SRC.
Similarly we might add that, as discussed in Ref.~\cite{Antonov:2011}, the effect of FSI on the shape of the missing momentum distribution is not unambiguously distinguishable from the effect of SRC.
These considerations, and other ambiguities related to e.g. the kinematic setup and the single-nucleon current~\cite{Rohe:2005,ONEILL1995}, mean that it is important to keep in mind the reference model used in the analysis when interpreting transparency data; this is not always straightforward.
RDWIA calculations were compared to the RMSGA model of Ref.~\cite{Ryckebusch:2003fc} and nuclear transparency data in Refs.~\cite{RDWIAvsRMSGA, Lava04}. 
Both models give similar results for $T_p > 200~\mathrm{MeV}$ and are consistent with the transparency data when SRC are taken into account. Hence they should be comparable to the NEUT cascade results presented in this work.
Measurements of the nuclear transparency are not available however in the region of lower $T_p$ where the rROP+NEUT, ROP, and RMSGA descriptions diverge from each other.

Measurements of the $(e,e'p)$ process at lower nucleon energy have been performed, 
but these do not provide the distribution of the outgoing nucleon's energy.
Measurements are performed for a fixed outgoing nucleon energy, for specific kinematics which minimize the effect of FSI, and restrictive cuts on the missing energy are applied in order to isolate the contribution from specific nuclear shells. 
The RDWIA approach with suitable optical potentials describes the shape of such data well~\cite{Udiaseep,Udias01}. 
The comparison of the model to data allows to extract a spectroscopic factor, i.e. a normalization factor which takes into account the depletion of a shell-model state. Again, the depletion of single-particle orbits due to SRC and due to FSI cannot be distinguished from each other just from one measurement.
It is the constancy of spectroscopic factors for a specific shell, measured at different outgoing nucleon energies and lepton kinematics, that would imply a correct description of the reduction due to FSI in the exclusive cross section~\cite{Dickhoff:2004}. 
This was found to be the case e.g. for the description of $(e,e'p)$ on oxygen with a similar RDWIA approach as used here in Refs.~\cite{Meucci:2001eep,Radici:2003zz}. While this constitutes a good indication of the reliability of the approach, measurements at a number of specific kinematics do not provide a global view of the effects of FSI.

\begin{figure*}
    \centering
    \includegraphics[width=0.45\textwidth]{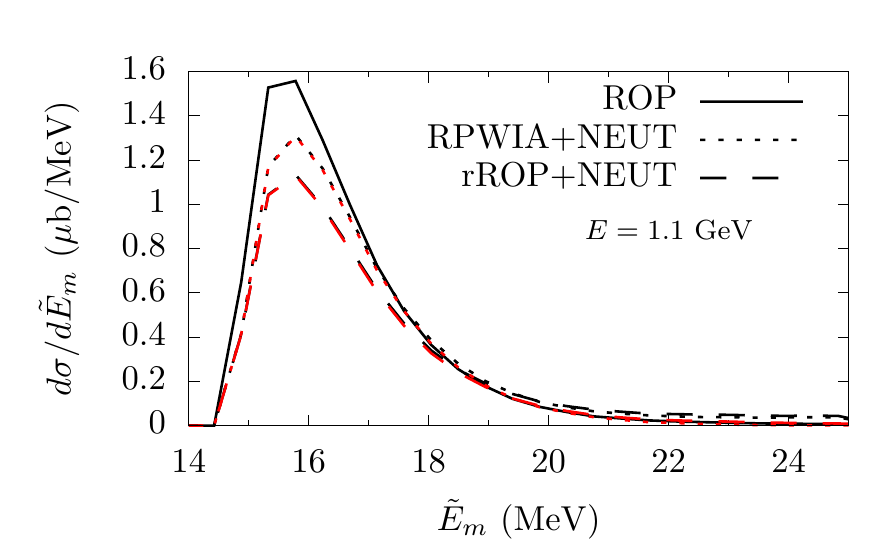} 
    \includegraphics[width=0.45\textwidth]{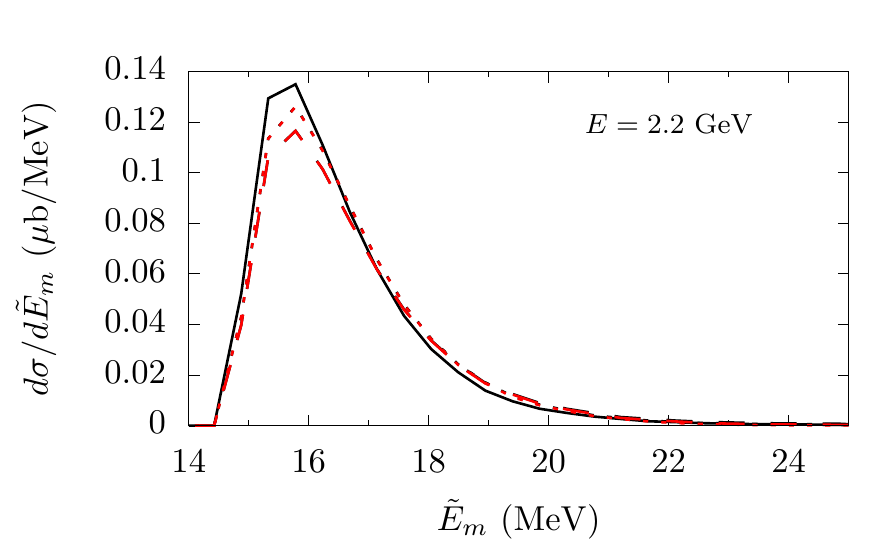} \\ 
    \includegraphics[width=0.45\textwidth]{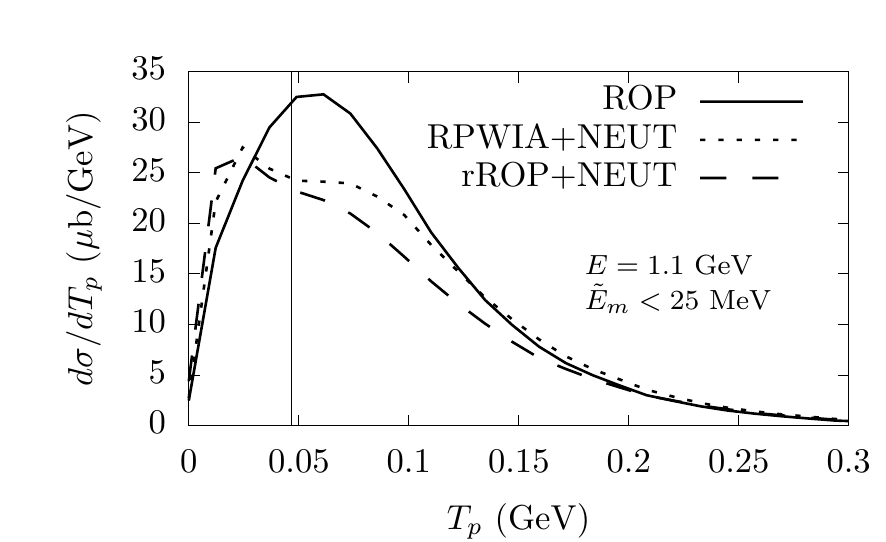} 
    \includegraphics[width=0.45\textwidth]{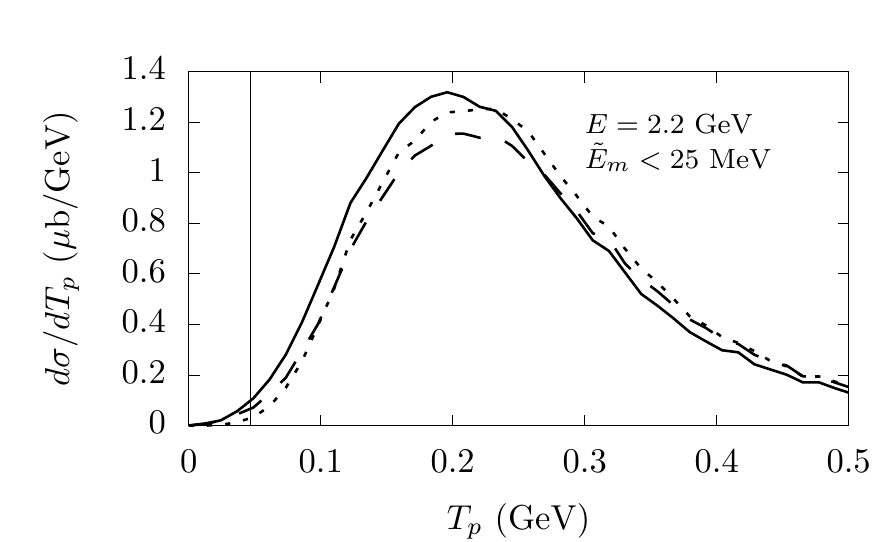} 
    \caption{Top panels show the $\tilde{E}_m$ distribution obtained for different models. The full results obtained in NEUT (black dashed and dotted lines) match the results in which only 1-track events are retained (corresponding red lines) closely.
    The bottom panels show the the proton kinetic energy distributions where $\tilde{E}_{m} < 25~\mathrm{MeV}$. Left(right) panels are for an incoming electron energy of $1.1(2.2)~\mathrm{GeV}$. Kinematic cuts applicable to the $e4\nu$ analysis of CLAS data (see text) are applied in all cases. The vertical line shows the threshold for proton detection ($p_p > 300~\mathrm{MeV}$).}
    \label{fig:eep}
\end{figure*}

In recent years, several collaborations have proposed to analyze existing electron scattering data, or even perform new measurements, with the express purpose of informing the modeling in neutrino-scattering experiments.
The $e4\nu$ collaboration uses data taken with the CLAS detector at Jefferson lab for this purpose, and has recently analysed $(e,e'p)$ data to test energy estimators used in neutrino experiments~\cite{CLAS:2021neh}. 
The open trigger and large angular acceptance of CLAS for charged particles, combined with a rather small threshold of $p_p > 300$ MeV for proton detection, makes this a rather dependable proxy for a neutrino experiment.
The fixed incoming energy means that complications due to flux-folding, as one has in a neutrino experiment, are not present.
However, when no additional restriction on the nuclear phase space is imposed one still faces the problem of multiple interaction mechanisms that are difficult to disentangle, in addition to the uncertainties in the description of the nucleus and FSI.

The combination of a restrictive cut on the missing energy, while still allowing a large range of kinematics for the outgoing lepton and proton, may directly inform the modeling of FSI relevant to neutrino scattering experiments. Such data might give insight in the differences between the rROP+NEUT cascade results and the ROP shown in this work, and the differences between different event-generators~\cite{MINERvA:2019neutrons, ValidationmethodsINC:2021} found in the low-$T_N$ region.

In Fig.~\ref{fig:eep} we show cross sections for $(e,e'p)$ on carbon.
The results are computed for fixed incoming energies, and are integrated over lepton energy and angles such that $15^\circ < \theta_{e^\prime} < 40^\circ$ and $E_{e^\prime} > 400~\mathrm{MeV}$. The proton scattering angle is restricted to $20^\circ < \theta_p < 140^\circ$, these cuts are suitable for an $e4\nu$ analysis of CLAS data~\cite{private:AdiA,CLAS:2021neh}. 
We apply a cut in missing-energy as defined in Eq.~(\ref{eq:Emiss}), $\tilde{E}_{m} < 25~\mathrm{MeV}$ which fully includes the $p$-shell region, and is below the threshold for two-nucleon knockout.
This simplified missing energy $\tilde{E}_m$, in which nuclear recoil is neglected, does not incorporate angular information such that it can be determined easier in experiments on an event-by-event basis. 
The cut in missing energy makes the contribution from multi-nucleon knockout and from inelastic FSI negligible.
The latter is shown in the upper panels of Fig.~\ref{fig:eep}, the red lines correspond to the $1$-track results, which practically match the full results (the corresponding black lines) in this kinematic region.
This means that inelastic FSI mostly leads to larger energy losses of the leading proton, and these events are not present for $\tilde{E}_{m} < 25~\mathrm{MeV}$. Albeit for the low energy cross section in the right panel, a small amount of additional strength is found in the high-$\tilde{E}_m$ tail compared to the $1$-track results. 
This cut in missing energy hence allows to study the treatment of FSI, without having to deal with additional confounding factors. 
Effects beyond the mean field, due to e.g. short-range correlations, would lead to a spreading of the missing energy profile and a reduced occupation of the $p$-shell.
Hence a spectroscopic factor  that takes into account the partial occupancy of the $p$-shell should be taken into account, this factor should be the same for all considered models. A shape comparison to experimental $T_p$ distributions can thus be performed straightforwardly, normalizing for example to the number of events for high $T_p$.

The results in the bottom panels of Fig.~\ref{fig:eep} show that discrepancies between the different treatments found for small $T_p$, are accessible mostly for lower incoming energies ($E_e = 1.1~\mathrm{GeV}$ in this case). 
The electromagnetic cross section is dominated by events at the most forward lepton scattering angle, which is $15^\circ$ in this case. Lower $T_p$ results at larger incoming energies would become more prominent if this threshold can be made even smaller.
We find a rather good agreement between the different models at high $T_p$.
Differences between the rROP+NEUT and ROP are more significant in the electromagnetic cross sections shown in Fig.~\ref{fig:eep} than for flux-averaged neutrino cross sections computed over the full lepton and hadron phase space, shown e.g. in Fig.~\ref{fig:NEUTshells}.
This follows mainly from the $1/(Q^2)^2$ weighting of the cross section combined with the restriction of the phase space to $\theta_{e^\prime} > 15^{\circ}$, it is hence important to consider the exact kinematic conditions when comparing to electron scattering data.
The confrontation of the results in this work, and those found in other cascade models, with electron scattering data  employing a restriction on $\tilde{E}_m$ should provide constraints on the modeling of FSI in cascade models.

Finally we note that while the cut on $\tilde{E}_m$ applied here might seem restrictive when considering the scope of a neutrino experiment, it is a necessary step to be able to distinguish and validate models for the (presumably) simplest interaction mechanism before tackling the severely more complicated unrestricted case.

\section{Conclusions \label{sec:conclusion}}
We have compared the treatment of FSI in the NEUT cascade model to the relativistic distorted wave impulse approximation (RDWIA) with  a relativistic optical potential (ROP) obtained from elastic $p-A$ scattering.
We have considered the single-nucleon knockout process, and provide results for the T2K near detector $\nu_\mu$-flux. 
As input to the cascade we used events generated from a RDWIA calculation in which only the real part of the optical potential is used (rROP). As discussed in Refs.~\cite{Gonzalez-Jimenez:2019ejf,Gonzalez-Jimenez19}, the rROP approach provides a robust description of  the inclusive cross section.
As the cascade model only affects the composition of the hadron final state, this inclusive cross section is recovered upon summation over all final-state topologies that follow from the cascade. Additionally, the present approach retains the relation between the hadron and lepton kinematics fully. This contrasts with factorized approaches that are commonly used in neutrino event generators which include only information on the inclusive cross section~\cite{Dolan:2021rdd,Dolan:2019bxf}.

In order to compare the ROP approach, which removes the number of nucleons that undergo inelastic interactions in the nucleus, with the cascade model which explicitly redistributes these nucleons over different final-state channels we introduced a cut on the missing energy.
The events from the NEUT cascade which pass the missing energy cut are comparable to a selection of events for which no interaction happens in the cascade.
With this cut, the energy distribution of leading protons from the NEUT cascade yields a cross section of the same shape and magnitude as the one found with the full optical potential for proton kinetic energies larger than $150$ MeV.
This result depends on the nuclear density used in the cascade model, for which we consistently use the density obtained in the RMF model used for the ROP calculations. The agreement between cascade and ROP disappears when a simple uniform density is used.
We performed this analysis for carbon, oxygen, and calcium nuclei and find that the cascade model yields similar results for all three, granted that the events and nuclear density used as input to the cascade are consistently obtained from the respective rROP calculations.
Such agreement provides a satisfying picture of the process as the rROP accounts for the modification of the nucleon's dispersion relation in the medium, while this effect is not included in the NEUT cascade. It is introduced here in the distribution of events to which the cascade is applied. 
While the comparison of ROP and cascade is robust for high energies, significant differences in shape and magnitude of more than 50 percent are found for proton kinetic energies below $100~\mathrm{MeV}$. This is likely related to the proton-nucleon reaction cross section being underpredicted by the cascade at low kinetic energy.
Additionally we find that when nucleons with momenta below the Fermi momentum are introduced in the NEUT cascade they always leave the nucleus without re-interaction, the ROP however predicts a much smaller cross section in this region compared to the events used as input. 
While the present study is limited to only the NEUT cascade, the results show that one should be skeptical towards semi-classical cascade models in the region of low nucleon energies. 

We compared the rROP+NEUT treatment to the data in terms of single-transverse variables obtained in T2K~\cite{T2K:2018rnz}. 
In this comparison the non-QE results of Ref.~\cite{Bourguille:2020bvw} were added.
We find reasonable results which seem mostly consistent with the data. While clear differences are found between the rROP+NEUT and pure ROP treatments, the comparison does not allow to clearly isolate the modeling of FSI because of the large contribution of interaction mechanisms beyond quasielastic one-nucleon knockout.
We discuss the prospect of resolving the discrepancies found at low to intermediate energies with electron scattering data, and provide results for kinematics applicable to the CLAS data for which the $e4\nu$ collaboration performs analyses.

This work outlines a method to apply the analysis of proton-nucleus elastic scattering to provide novel constraints for cascade models.
Comparisons of other cascade models to the ROP calculations employed here can provide additional insight and possible validation of the treatment of FSI in neutrino event generators.
The event distributions that where computed for this work are available for such studies upon reasonable request.
Similar comparisons, with a more intricate spectral function, are currently being performed with the NuWro event generator~\cite{NuWro:FSI,Niewczas:prep}.
Finally we note that the events used as input to the cascade model in this work are more sophisticated than those generally used in neutrino experiments. In the latter, the hadron and lepton information are often factorised as discussed, e.g., in Refs~\cite{Dolan:2019bxf}, while in this case an unfactorized calculation which treats FSI in a quantum-mechanical manner is used. In this respect the events used in this work can be used to estimate the uncertainty of, or to improve on such factorized approaches.


\acknowledgments
A.N. thanks the DPNC at the University of Geneva where part of this work was performed for their hospitality.
This research was supported by the Research Foundation Flanders (FWO-Flanders); the Spanish government through RTI2018-098868-B-100; the government of Madrid and Complutense University under project PR65/19-22430 (R.G.-J); the Swiss National Foundation through grant No.~200021\_85012; the Polish Ministry of Science and Higher Education under grant No.~DIR/WK/2017/05, and the NCN grant No.~2020/37/N/ST2/01751 (K.N.); the Swiss confederation through ESKAS nr.~2020.0004; This manuscript has been authored by Fermi Research Alliance, LLC under Contract No.~DE-AC02-07CH11359 with the U.S. Department of Energy, Office of Science, Office of High Energy Physics.


\bibliographystyle{apsrev4-1.bst}
\bibliography{BiblioG}

\end{document}